\documentclass{segabs}
\usepackage[tight,footnotesize]{subfigure}
\usepackage{amsmath} % include the amstext, amsbsy, and amsopn packages
\usepackage{amssymb,stmaryrd} % extra symbols
\usepackage{graphicx}
\usepackage{color}
\usepackage{multirow}% http://ctan.org/pkg/multirow
\usepackage{hhline}% http://ctan.org/pkg/hhline

\renewcommand{\figdir}{Fig} % figure directory

\def\RB{{\mathbb R}}

\begin{document}

\title{Spatial-Temporal Densely Connected Convolutional Networks: An Application to CO$_2$ Leakage Detection}

\renewcommand{\thefootnote}{\fnsymbol{footnote}} 

\author{Zheng Zhou\footnotemark[1], Youzuo Lin,  Yue Wu, Los Alamos National Laboratory; Zan Wang, Robert Dilmore,  National Energy Technology Laboratory; and George Guthrie, Los Alamos National Laboratory}

\righthead{Spatial-Temporal Densely Connected Convolutional Networks: An Application to CO$_2$ Leakage Detection}

\maketitle

\begin{abstract}
In carbon capture and sequestration, building an effective monitoring method is a crucial step to detect and respond to CO$_2$ leakage.  CO$_2$ leakage detection methods rely on geophysical observations and monitoring sensor network. However, traditional methods usually require physical models to be interpreted by experts, and the accuracy of these methods will be restricted by different application conditions. In this paper, we develop a novel data-driven detection method based on densely connected convolutional networks. Our detection method learns a mapping relation between seismic data and the CO$_2$ leakage mass. To account for the spatial and temporal characteristics of seismic data, we design a novel network architecture by combining 1-D and 2-D convolutional neural networks together. To overcome the expensive computational cost, we further apply a densely-connecting policy to our network architecture to reduce the network parameters.  We employ our detection method to synthetic seismic datasets using Kimberlina model. The numerical results show that our leakage detection method accurately detects the leakage mass. Therefore, our novel CO$_2$ leakage detection method has great potential for monitoring CO$_2$ storage.
\end{abstract}

\section*{Introduction}
The carbon capture and sequestration (CCS) technology collects the CO$_2$ from industrial sources such as thermal power plant and then injects compressed CO$_2$ into certain geologic formations underground. As the geological structure gradually changes, the sequestered CO$_2$ may leak into the atmosphere or mingle with the underground drinking water layer. Once the leakage rate exceeds the safety level, it will become a threat to the environment and public health~\citep{Yang-Probabilistic-2011}. To resolve this problem, several monitoring technologies have been developed to detect CO$_2$ leakage at sequestration sites including observation of artificial seismic data, groundwater chemistry monitoring, near-surface measurements of soil CO$_2$ fluxes, analysis of carbon isotopes in soil gas, measurement of tracer compounds injected with the sequestered CO$_2$, and nearby atmospheric monitoring of CO$_2$ and tracer gases~\citep{Quantification-2011-Korre, Leuning-Atmospheric-2008, Benson-Monitoring-2007}. Among all these techniques, monitoring leakage through seismic data is the most powerful in terms of plume mapping, quantification of the injected volume in the reservoir and early detection
of leakage~\citep{Geophysical-2011-Fabriol}.  Many related detection methods have sprung up in this field, such as obtaining the elastic parameters at different injection times through Gassmann fluid substitution to estimate the CO$_2$ sequestration status~\citep{Macquet-Feasibility-2017}.

With rapid improvements in computational power and fast data storage, machine learning techniques have been effectively applied to problems from various domains. Deep learning, a technique with its foundation in artificial neural networks, is emerging in recent years as a powerful tool~\citep{Deep-2015-LeCun}. Among various deep learning methods, convolutional neural networks~(CNN) have achieved promising results in both detection and prediction tasks, such as speech recognition in one-dimensional voice signal~\citep{Abdel-Convolutional-2014} and semantic segmentation in two-dimensional image data~\citep{Jonathan-Fully-2015}. 

\begin{figure*}[h]
	\begin{center}
		\centering
		\includegraphics[width=1.1\columnwidth]{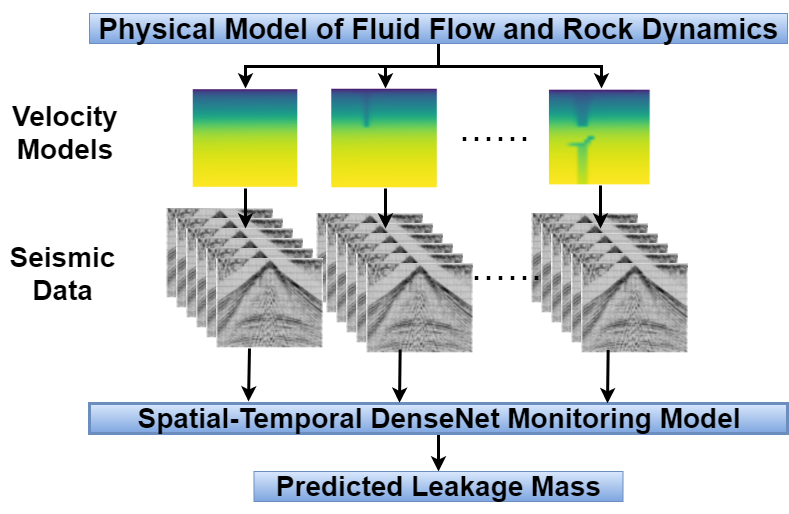}
	\end{center}
	\caption{The schematic illustration of our detection method. In the data generation stage, the velocity models are generated from the fluid and rock simulations, and the simulated seismic data are further obtained from these velocity models. Next stage is model training. Majority of seismic data along with actual CO$_2$ leakage mass are fit into our ST-DenseNet model to learn an appropriate mapping relation between seismic data and leakage mass. The rest of seismic data are then fit into our fine-trained model to validate its performance.}
    \label{fig:ST-DensNet}
\end{figure*}

In this paper, we develop a novel end-to-end data-driven detection method, which directly learns the mapping relation from seismic data to CO$_2$ leakage mass as shown in Figure~\ref{fig:ST-DensNet}. We design our detection model based on conventional CNN architecture. Seismic data comes with typical the spatial and temporal characterics. The seismic trace from the single receiver is a typical 1-D time series, while the 2-D seismogram collected from multiple receivers can be treated as imagery, and there is spatial relevance between different traces. Therefore, instead of simply adapting the existing CNN architectures, we design a novel network architecture by combining 1-D and 2-D CNN together to account for both spatial- and temporal characteristics of seismic data.  Training CNNs can be computationally expensive, not to mention a combination of a couple of CNNs. To overcome the expensive computational cost, we further apply a densely-connecting policy to our architecture to reduce the parameters. 

We validate the performance of our detection method using synthetic seismic datasets generated based on Kimberlina model~\citep{Simulated-2017-Buscheck}. Our detection method yields more accurate detection of CO$_2$ leakage mass by comparing to other machine learning/deep learning techniques. 

%Our seismic data is collected from 100 sensors evenly arranged on the surface and last for 6,000 time stamps in total. 

\section*{Methodology}

%\multiplot{1}{LearningProcedure-Measurements-cut}{width=0.75\columnwidth}
%{a}

%---------------------------------Figure---------------------------------%
%---------------------------------Figure---------------------------------%

\subsection{Convolutional Neural Network (CNN)}
The convolutional neural network is one of the most influential network structures in deep learning. LeNet~\citep{lecun1995learning}, which is known as the first kind of CNN, is proposed by Yann LeCun in 1995. In 2012, AlexNet~\citep{krizhevsky2012imagenet} won the ImageNet champion. The authors introduced fully connected layers and max-pooling layers to help AlexNet outperform all the other methods. After that, a sequence of different structures such as VGG~\citep{VGG}, ResNet~\citep{he2016deep}, GoogleNet~\citep{szegedy2017inception}, and DenseNet~\citep{huang2017densely} sprung up.  

Provided with $\hat{Y} \in \RB^{1}$ as predicted leakage mass, $Y\in \RB^{1}$ as groundtruth. We utilize the Mean Square Error (MSE) as the loss function to measure the distance between groundtruth and predicted value
\begin{align}
Loss_{MSE} = \frac{1}{n}\sum^{n}_{i=1}(Y_i-\hat{Y_i})^2,
\end{align}
where $n$ is the sample size. 

\subsection{Residual Network (ResNet)}
Consider a single image $x_0$ that is passed through a convolutional network. The network comprises $L$ layers, each of which implements a non-linear transformation $H_l(\cdot)$,
where $l$ indexes the layer. $H_l(\cdot)$ can be a composite function of operations such as batch normalization (BN), rectified linear units (ReLU), pooling, or convolution. We denote the output of the $l^{\mathrm{th}}$ layer as $x_l$.

Traditional convolutional feed-forward networks connect the output of the $l^{\mathrm{th}}$ layer as input to the $(l+1)^{\mathrm{th}}$ layer, which gives rise to the following layer transition: $x_l = H_l(x_{l-1})$. As shown in Figure~\ref{fig:res_dense}(a), ResNet adds a skip-connection that bypasses the non-linear transformations with an identity function~\citep{he2016deep}
\begin{align}
x_l = H_l(x_{l-1})+x_{l-1}.
\label{eq:resblock}
\end{align}

\subsection{Densely Connected Network (DenseNet)}
The intuition behind densely connected networks is similar to ResNet. Both of them aim at reusing the convolution features from previous layers and reducing the number of trainable parameters.

\begin{figure}[h]
\centering
\subfigure[]{\includegraphics[width=0.3\linewidth]{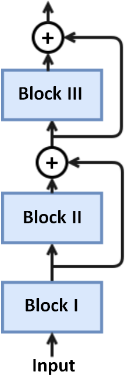}}
\hfil
\subfigure[]{\includegraphics[width=0.3\linewidth]{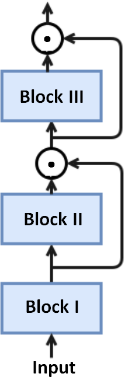}}
	\caption{A schematic illustration of the ResNet (a) and DensNet (b), where ``$\bigoplus$'' represents feature addition operation and ``$\bigodot$'' stands for feature concatenation operation.}
	\label{fig:res_dense}
\end{figure}

As shown in Figure~\ref{fig:res_dense}(b), a densely connected block is formulated as
\begin{align}
x_{l + 1} & = \mathcal{H}([x_{0}, x_{1}, ..., x_{l}]),\\
\mathcal{H}(x) & = W*(\sigma(B(x))),
\label{eq:denseblock}
\end{align}
where $W$ is the weight matrix, the operator of ``*'' denotes convolution, $B$ denotes batch normalization (BN), $\sigma(x) = \text{max}(0, x)$ and $[x_{0}, x_{1}, ..., x_{l}]$ denotes the concatenation of all outputs of previous layers. 

%Note that
%$
%K \; \approx \; C W^\dag C \; = \; \Psi \Psi^T .
%$

%---------------------------------Figure---------------------------------%
% \begin{figure}[h]
% 	\begin{center}
% 		\centering
% 		\includegraphics[width=0.75\columnwidth]{Fig/res_dense}
% 	\end{center}
% 	\caption{A schematic illustration of the ResNet and DensNet. Where '$+$' in round node represents feature addition operation, and '$\bullet$' stands for feature concatenation operation.}
% 	\label{fig:res_dense}
% \end{figure}

\subsection{Spatial-Temporal DenseNet (ST-DenseNet)}
The seismic trace from the single receiver is a typical 1-D time series, while the 2-D seismogram collected from multiple receivers can be treated as imagery, and there is spatial relevance between different traces. In order to account for both the spatial and temporal characteristics, we develop a new network structure, called ``spatial-temporal DenseNet (ST-DenseNet)'', by combining 1-D and 2-D CNNs as shown in Tabel~\ref{table:densenet_architect}. The major difference between conventional DenseNet and our ST-DenseNet is that our model starts with applying convolution layers to 1-D time series, and follows by employing convolutions on 2-D seismograms. 

There are a couple of benefits of using our ST-DenseNet. Firstly, applying 1-D convolution layers in the time domain can not only reduce the size of model parameters but also learn important temporal features. Secondly, applying 2-D convolution layers will fuse high-level spatial and temporal features together. All these benefits turn out to be critical in improving prediction accuracy and reduce training costs.

% could take the best use of these temporal information, so that our network is able to focus on time domain at the first stage. Secondly, the following 2-D convolution layers will merge high level spatial and temporal features together. Thirdly,  starting our network with 1-D convolution is that the number of temporal features could be rapidly reduced. 

% Distinguish from image data, two dimensions of spatial-temporal data have different physical meanings. Directly operating them as 2-D image data is not a wise choice, especially while temporal dimension consists much detailed information (such as our seismic data). 

% Applying 1-D convolution in time domain could take the best use of these temporal information, so that our network is able to focus on time domain at the first stage, and the following 2-D convolution layers will merge high level spatial and temporal features together. Another advantage of starting our network with 1-D convolution is that the number of temporal features could be rapidly reduced. As we can see in Table~\ref{table:results1}， our Spatial-Temporal DenseNet has the least parameters. 

\begin {table}
\centering
\begin{tabular}{ |c|c|c| }
\hline
Stage & Layers & Dim.\\
\hline
Input & - & 6000 $\times$ 100 $\times$ 6 \\
\hline
Conv1D-1 & conv(7$\times$1), 32, /(4$\times$1) & 1499 $\times$ 100$\times$32 \\
\hline
Conv1D-2 & conv(5$\times$1), 32, /(3$\times$1) & 499 $\times$ 100 $\times$ 32 \\
\hline
Pool1D & max-pool(2$\times$1), /(2$\times$1) & 249 $\times$ 100$\times$32 \\
\hline
Conv1D-3 & conv(3$\times$1), 32, /(2$\times$1)) & 124 $\times$ 100 $\times$ 32 \\
\hline
Dense2D-1 & [conv(3$\times$3), 64] $\times$ 3 & 124 $\times$ 100 $\times$ 224 \\
\hline
\multirow{2}{*}{Transition} & conv(1$\times$1), 64 & 124 $\times$ 100 $\times$ 64 \\
\hhline{~--} & max-pool(2$\times$2), /(2$\times$2) & 62 $\times$ 50 $\times$ 64 \\
\hline
Dense2D-2 & [conv(3$\times$3), 128] $\times$ 3 & 62 $\times$ 50 $\times$ 448 \\
\hline
\multirow{2}{*}{Transition} & conv(1$\times$1), 128 & 62 $\times$ 50 $\times$ 128 \\
\hhline{~--} & max-pool(2$\times$2), /(2$\times$2) & 31 $\times$ 25 $\times$ 128 \\
\hline
Dense2D-3 & [conv(3$\times$3), 256] $\times$ 3 & 31 $\times$ 25 $\times$ 896 \\
\hline
\multirow{2}{*}{Transition} & conv(1$\times$1), 256 & 31 $\times$ 25 $\times$ 256 \\
\hhline{~--} & max-pool(2$\times$2), /(2$\times$2) & 15 $\times$ 12 $\times$ 256 \\
\hline
Dense2D-4 & [conv(3$\times$3), 512] $\times$ 3 & 15 $\times$ 12 $\times$ 1792 \\
\hline
\multirow{2}{*}{Transition} & conv(1$\times$1), 512 & 15 $\times$ 12 $\times$ 512 \\
\hhline{~--} & max-pool(2$\times$2), /(2$\times$2) & 7 $\times$ 6 $\times$ 512 \\
\hline
Dense2D-5 & [conv(3$\times$3), 1024] $\times$ 3 & 7 $\times$ 6 $\times$ 3584 \\
\hline
\multirow{2}{*}{Transition} & conv(1$\times$1), 1024 & 7 $\times$ 6 $\times$ 1024 \\
\hhline{~--} & max-pool(2$\times$2), /(2$\times$2) & 3 $\times$ 3 $\times$ 1024 \\
\hline
\multicolumn{3}{|c|}{Flatten Layer} \\
\hline
\multicolumn{3}{|c|}{1-d fully connected, mean squared error loss} \\
\hline
\end{tabular}
\caption{Network architecture of our ST-DenseNet. This model is designed for inputs with 6,000$\times$100$\times$6, which are the simulated seismic data. ''Conv(7$\times$1), 32, /(4$\times$1)'' denotes using 32 (7$\times$1) convolution kernels with stride (4$\times$1).}
\label{table:densenet_architect}
\end {table}

\section*{Numerical Results}
\subsection{Dataset}
To evaluate the effectiveness of our proposed approach, we test and validate our model for CO$_2$ leakage mass prediction task on a simulated seismic dataset. The simulations were generated from a model framework based on a hypothetical, compartmentalized, CO$_2$ storage reservoir in the Vedder Fm. near Kimberlina in the southern San Joaquin Basin, California \citep{Simulated-2017-Buscheck}. A total of $2,927$ groups of simulations with different leakage mass are generated and we select 2,400 groups for training, 300 groups for validating, and 227 groups for testing. To generate the seismic data, a total of 3 sources and 100 receivers are evenly distributed along the top boundary of the model. The source interval is $500$~m, and the receiver interval is $15$~m. We use a Ricker wavelet with a center frequency of $50$~Hz as the source time function and a staggered-grid finite-difference scheme with a perfectly matched layered absorbing boundary condition to generate synthetic seismic reflection data. The synthetic trace at each receiver is a collection of time-series data of length $6,000$. We employ our new prediction method to estimate the CO$_2$ leakage mass directly from the seismic data. As for the computing environment, we run our tests on a computer with Intel Xeon E5-2650 core running at 2.3~GHz, and Tesla K40c GPU with 875~MHz boost clock.

\subsection{Data Preprocessing}
% Before we start training our model, we combine the seismic data which measure the same velocity model together. Since we have three different seismic wave generating sources, and the seismic wave produced from the same source have two different components (the seismic wave could vibrate in horizontal direction X and in vertical direction Z), each group of seismic which corresponding to the same velocity model would have 6 channels, and the size of each channel remains as 6,000 $\times$ 100.

% While building our ResNet-based and DensNet-based regression model from the original CNN networks, we create the residual connection (feature addition) and dense connection (feature concatenation) between each 2-D convolution blocks. 
Considering the order of CO$_2$ leakage mass can vary from $0$ to $10^{10}$~(tonne), all the mass data are standardized by Log function
\begin{align}
\tilde{Y} = \log_{10}(Y+1), 
\end{align}
where $Y$ represents the original value of CO$_2$ leakage mass, and $\tilde{Y}$ stands for standardized leakage mass. Each $Y$ is added by 1 to avoid taking Log at 0.
% We also would like to further reduce the number of parameters and improve the generation of our model, in implementation of our experiments we use a 3-D convolution kernel to replace the flatten layer and 1-D fully connected layer.

\subsection{Test: CO$_2$ Leakage Mass Regression}

\begin{table*}[ht]
\centering
\begin{tabular}{ |c|c|c|c|c| }
\hline
& $\pm10\%$ Acc & $\pm5\%$Acc & $\pm3\%$Acc & Parameters \\ 
\hline
Kernel SVR & 0.316 & fail & fail & 36K \\
\hline
VGG-based CNN& 0.895 & 0.864 & 0.821 & 29M \\
\hline
ResNet-based CNN& \textbf{0.988} & 0.937 & 0.906 & 17M \\
\hline
ST-DenseNet& 0.983 & \textbf{0.958} & \textbf{0.913} & 9M \\
\hline
\end{tabular}
\caption{The CO$_2$ leakage mass prediction results given by kernel support vector regression (kernel SVR), VGG-based CNN, ResNet-based CNN and our ST-DenseNet. ``fail'' indicates the accuracy is extremely low. The results indicate that our ST-DenseNet outperforms both the classical regression model and other CNN-based models.}
\label{table:results1}
\end{table*}

We evaluate the performance of our ST-DenseNet detection method on CO$_2$ leakage mass prediction task. We compare our method to different regression models including 1. support vector regression with radial basis function kernel~(Kernel SVR); 2. VGG-based~CNN; and 3. ResNet-based CNN. Multiple metrics are used to evaluate the regression results. In particular, ``$\pm r\%$ Accuracy'' represents the tolerance of error rate. In other words, if the predicted leakage mass is in the range from (100-r)\% to (100+r)\% of the actual leakage mass, the prediction result is considered as accurate. Otherwise, the prediction fails. In our test, we select three different values of $r=3, 5$ and $10$. The number of trainable parameters is also reported to reflect the computational complexity of different models.

The results from several different methods are provided in Table~\ref{table:results1}. We notice that the kernel SVM has extremely low accuracy. So, we put ``fail'' in the table. It suggests that advanced feature extraction techniques are required in advance to apply SVR or other classical regression models. By comparing to the VGG-based CNN or ResNet-based CNN, our ST-DenseNet yields higher accuracy. The only exception is the testing scenario of ``$\pm10\%$ Accuracy'', where our method still produces comparable results to those obtained by using ResNet-based CNN. The number of trainable parameters can be used as an indication of the computation cost. In Table~\ref{table:results1}, we observe that our ST-DenseNet yields the smallest number of model parameters among all three CNN-based methods. So, we conclude that our ST-DenseNet can not only produce the most accurate leakage mass prediction but also requires the least amount of trainable parameters. 
\begin{figure*}[ht]
\centerline{
\subfigure[]{\includegraphics[width=0.33\linewidth]{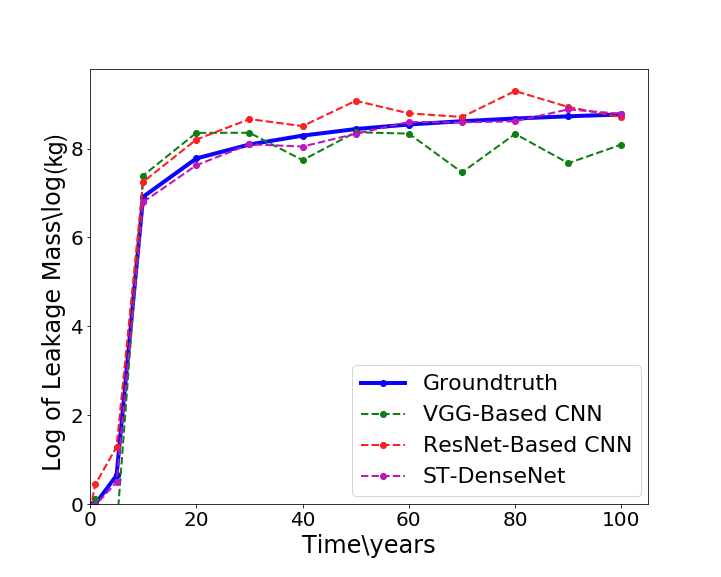}}
\subfigure[]{\includegraphics[width=0.33\linewidth]{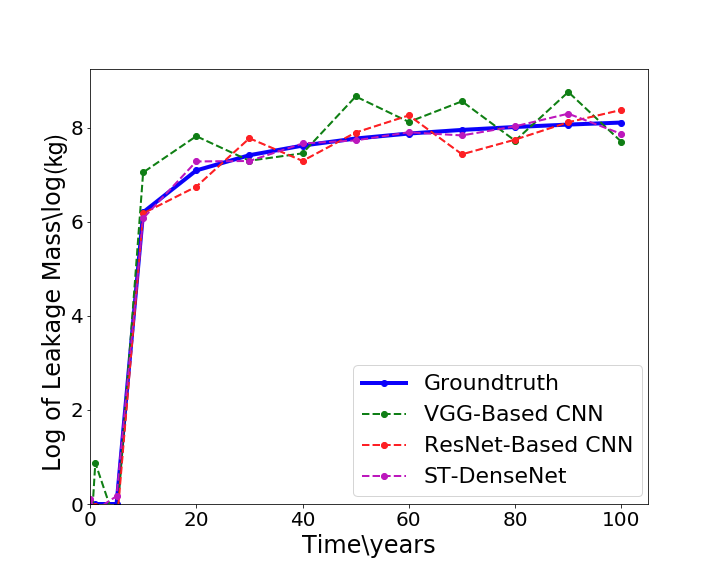}}}
\centerline{
\subfigure[]{\includegraphics[width=0.33\linewidth]{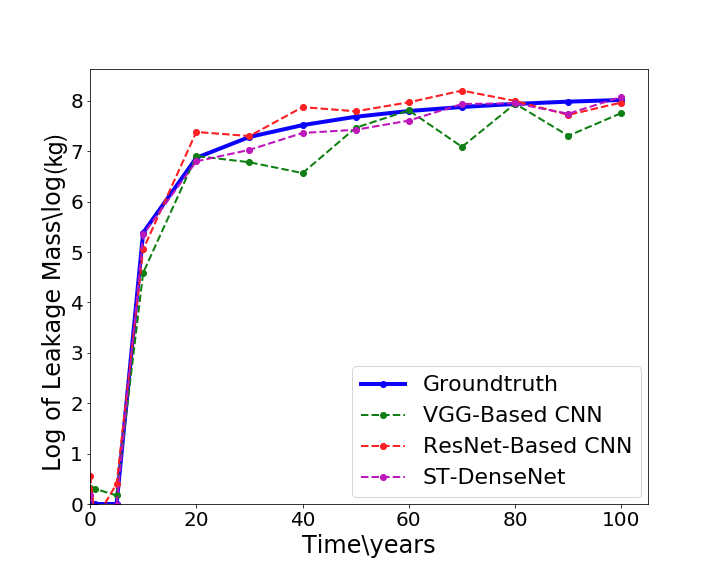}} 
\subfigure[]{\includegraphics[width=0.33\linewidth]{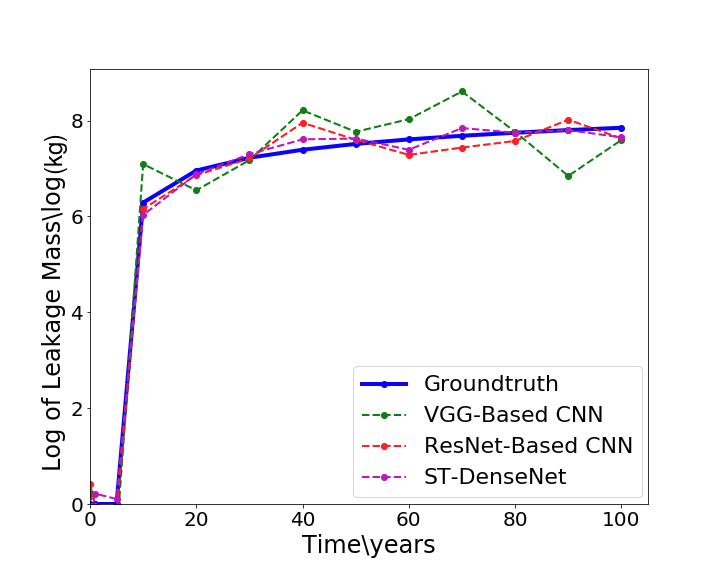}}
}
\caption{Illustration of four CO$_2$ leakage mass predictions. Each of these four results contains 13 different leakage mass. The groundtruth is plotted in blue. Results obtained using VGG-based CNN (in green), ResNet-based CNN (in red), and our ST-DenseNET (in purple) are all provided. Our ST-DenseNet achieves the most accurate CO$_2$ leakage mass prediction result among all three methods. }
\label{fig:exp_detection}
\end{figure*}

In Figure~\ref{fig:exp_detection}, we provide the prediction results of the leakage mass at every year using the aforementioned three CNN methods. Each of these four results contains 13 different leakage mass, which varies with respect to time. The groundtruth is plotted in blue. Results obtained using VGG-based CNN (in green), ResNet-based CNN (in red), and our ST-DenseNET (in purple) are all provided in Figure~\ref{fig:exp_detection}. We notice that the leakage rate of CO$_2$ always experiences a rapid growth between the 5th and the 20th year, and then it remains a less abrupt leakage rate for the rest of time. Comparing all three CNN models, VGG-based CNN not only yields the most un-stable leakage detection but also produces the largest MSE. We suspect that this is due to the lack of skip connection structure. ResNet-based CNN model yields better performance than VGG-based CNN model. Benefited from the temporal 1-D convolution layers and spatial 2-D convolution layers, our ST-DenseNet produces the most accurate leakage prediction. This is consistent with the results shown in Table~\ref{table:results1}.

\section*{Conclusions}

In this paper, we developed a novel spatial-temporal densely connected convolutional networks structure for CO$_2$ leakage mass detection using seismic data.
Our method not only accounts for both the spatial and temporal characteristics of seismic data but also significantly reduces the number of parameters in the network structure. We apply our method to detect CO$_2$ leakage mass detection based on simulated seismic data. By comparing with several commonly-used machine learning methods, we demonstrate that our detection method has outperformed traditional regression methods and other popular CNN models. Therefore, our novel detection method shows great performance in CO$_2$ leakage detection and have potential for various subsurface applications.

\section{ACKNOWLEDGMENTS}

This work was co-funded by the U.S. DOE Office of Fossil Energy’s Carbon Storage program and the Center for Space and Earth Science~(CSES) at Los Alamos National Laboratory (LANL). The computation was performed using super-computers of LANL's Institutional Computing Program.

\onecolumn

%\append{The source of the bibliography}
%\verbatiminput{reference_SEG.bib}

% \twocolumn

\bibliographystyle{seg}  % style file is seg.bst
\bibliography{reference_SEG}

\end{document}